# Distinguishing the EH$_1$ and S$_1$ defects in n-type 4*H*-SiC by Laplace DLTS

Tihomir Knežević[1], Tomislav Brodar[1], Vladimir Radulović[2], Luka Snoj[2], Takahiro Makino[3], and Ivana Capan[1]*

[1]*Ruđer Bošković Institute, Bijenička 54, 10 000 Zagreb, Croatia*
[2]*Jožef Stefan Institute, Jamova 31, 1000 Ljubljana, Slovenia*
[3]*National Institutes for Quantum Science and Technology, 1233 Watanuki, Takasaki*

**Abstract:** We report on the low-energy electron and fast neutron irradiated 4*H*-SiC studied by deep-level transient spectroscopy (DLTS) and Laplace DLTS. Irradiations introduced two defects, $E_c$-0.4eV and $E_c$-0.7eV. They were previously assigned to carbon interstitial (C$_i$) labeled as EH$_{1/3}$ and silicon-vacancy (V$_{Si}$) labeled as S$_{1/2}$, for the low-energy electron and fast neutron irradiation, respectively. This work demonstrates how Laplace DLTS can be used as a useful tool for distinguishing the EH$_1$ and S$_1$ defects. We show that EH$_1$ consists of a single emission line arising from the C$_i$(*h*), while S$_1$ has two emission lines arising from the V$_{Si}$(*h*) and V$_{Si}$(*k*) lattice sites.

Electrically active radiation-induced defects in 4*H*-SiC have been extensively studied for decades. A whole variety of electrically active defects introduced by different radiation sources such as protons, electrons, neutrons, and ions have been revealed. Among them, two electrically active defects with deep levels at $E_c$-0.4eV and $E_c$-0.7eV in n-type 4*H*-SiC material have captivated special attention in the SiC community. They are peculiar as they always occur together, and independently of the radiation source.[1-12] Not until recently, they were commonly assigned to the same defect. However, the origin of these two radiation-induced defects has finally been elucidated.

Alfieri and Mihaela[3] reported the isothermal annealing study of the low energy (116 keV) electron radiation-induced defects in n-type 4*H*-SiC. They demonstrated beyond doubt that EH$_1$ ($E_c$-0.4eV) and EH$_3$ ($E_c$-0.7eV) are two different charge states of the same defect, and whose origin is related to a carbon interstitial (C$_i$). Moreover, they highlighted that EH$_{1/3}$ are not the same as S$_1$ ($E_c$-0.4eV) and S$_2$ ($E_c$-0.4eV). The EH$_{1/3}$ radiation-induced defects are introduced exclusively in the case of low-energy electron irradiation (<200 keV), since the silicon atoms cannot be displaced under such conditions.[3]

Outstanding progress in understanding S$_1$ and S$_2$ defects in proton-irradiated 4*H*-SiC was made by Bathen et al.[12] They identified S$_1$ and S$_2$ as silicon-vacancy V$_{si}$ (−3/=) and V$_{Si}$ (=/−) charge state transitions by combining photoluminescence (PL), DLTS, and hybrid density functional theory (DFT) calculations.

Although the origin of the $EH_1/S_1$ ($E_c$-0.4eV) and $EH_3/S_2$ ($E_c$-0.7eV) defects has now been resolved, they still cause some confusion in the analysis of DLTS data. The reason for this is the following: DLTS signals arising from the $EH_1$ and $S_1$, as well as signals from the $EH_3$ and $S_2$, are identical in the DLTS spectra and cannot be resolved.

Since its first demonstration in 1974, DLTS has proven to be the most sensitive technique for measuring electrically active defects in semiconductors, as it can detect defects at concentrations of about $10^{10}$ cm$^{-3}$. [13,14] The main shortcoming lies in energy resolution. Two closely spaced deep energy levels cannot be resolved. However, Laplace DLTS, an isothermal technique that provides a spectral plot of a processed capacitance signal as a function of emission rate rather than temperature, gives an order of magnitude better energy resolution. [13,14]

Laplace DLTS has already proven to be a very useful technique for resolving the overlapping DLTS signals in 4$H$-SiC material. Here we will briefly describe the most significant examples. Alfieri and Kimoto[15] have successfully applied Laplace DLTS to resolve the overlapping emission rates of the $EH_6$ and $EH_7$, which constitute the $EH_{6/7}$ peak in the DLTS spectrum. Following the results presented by Hemingsson et al.[16], in their pioneering work on $Z_{1/2}$, Capan et al.[17] have provided conclusive evidence that the most dominant DLTS peak, known as $Z_{1/2}$, consists of two emission lines, $Z_1$ and $Z_2$. These are assigned to the -$h$ and -$k$ configurations of the carbon vacancy ($V_C$). [17, 18] A more recent example is the work of Bathen et al.[12] on proton irradiated 4$H$-SiC and $S_1/S_2$ defects. Using Laplace DLTS, they resolved two emission lines arising from the $S_1$ and assigned them to the $h$ and $k$ configurations of the $V_{si}(-3/=)$.

The main aim of this work is to demonstrate how Laplace DLTS can be used for distinguishing the $EH_1$ ($E_c$-0.4eV) and $S_1$ ($E_c$-0.4eV) defects in n-type 4$H$-SiC irradiated with low-energy electrons and fast neutrons.

We used n-type nitrogen-doped 4$H$-SiC epitaxial layers, 25 μm thick. The SiC epilayers were grown on an 8° off-cut silicon face of a 350 μm thick 4H-SiC (0001) wafer without a buffer layer.[19] The Schottky barrier diodes (SBDs) were formed by thermal evaporation of nickel through a metal mask (100 nm), while the Ohmic contacts were formed on the backside of the SiC substrate by nickel sintering at 950 °C in an Ar atmosphere.

All irradiations were performed after the SBDs fabrication, through the nickel contact and at room temperature (RT). Electron irradiation was performed at Nissin Electric Group (NEG), Kyoto, Japan. The electron energy was 150 keV, and the total fluence was $1\times10^{15}$ cm$^{-2}$. Neutron irradiation was performed at the Jozef Stefan Institute (JSI) TRIGA Mark II reactor, Ljubljana, Slovenia. The neutron fluence was $1\times10^{13}$ cm$^{-2}$. Thermal neutrons with energy less than 0.55 eV were filtered by a cadmium box with a wall thickness of 1 mm. More details on neutron irradiation are given elsewhere.[20, 21]

Figure 1 shows an as-grown 4H-SiC sample. As expected, only one DLTS peak is detected in the as-grown n-type 4$H$-SiC material. This is a widely known $Z_{1/2}$, with the activation energy for electron emission of 0.67 eV. It is the scrutinized defect in 4$H$-SiC, earlier assigned to the carbon vacancy ($V_C$).[22,23]

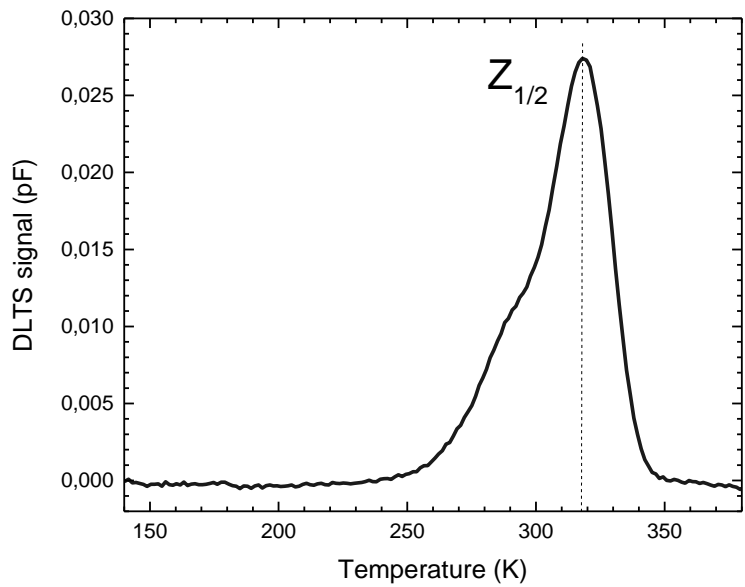

Figure 1. DLTS spectrum for the as-grown 4*H*-SiC sample.

Figure 2 shows DLTS spectra for n-type 4*H*-SiC samples irradiated with low-energy electrons and fast neutrons.

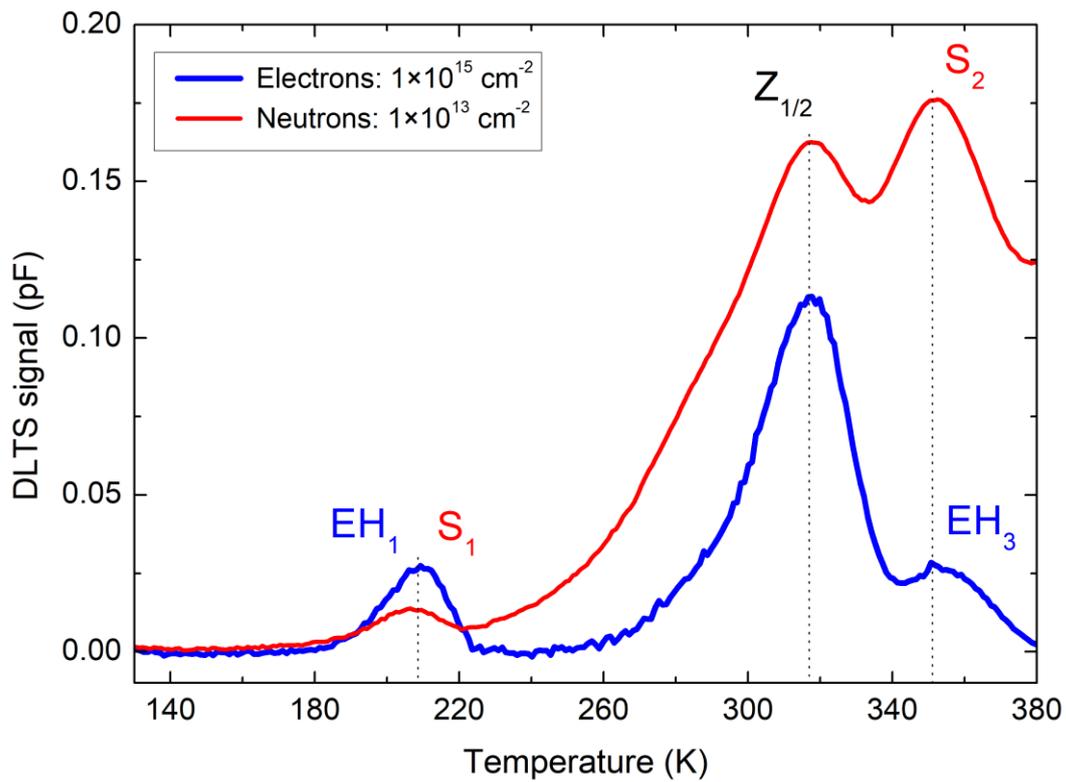

Figure 2. DLTS spectra for the 4H-SiC samples irradiated with low-energy electrons and neutrons.

In addition to $Z_{1/2}$, two deep level defects at $E_c$-0.4eV (EH$_1$/S$_1$) and $E_c$-0.7eV (EH$_3$/S$_2$) are detected in irradiated 4*H*-SiC samples. As previously described in the text, these traps were assigned to the Ci-related defects (EH$_1$ in the low energy electron irradiated samples) and V$_{si}$ (S$_1$ in the fast neutron irradiated samples). An interesting feature has been observed in Figure 2. While the ratio between EH$_1$:EH$_3$ is 1:1, introduction rates for S$_1$ and S$_2$ significantly differ.

This brings us to the main objective of this work. How can we distinguish the $E_c$-0.4eV level in the low energy electron (EH$_1$) or fast neutron (S$_1$) irradiated 4*H*-SiC samples? Following the examples of successful application of Laplace DLTS technique, we have applied Laplace DLTS to make further progress in understanding EH$_1$ and S$_1$ defects.

Figure 3 shows Laplace DLTS measurements at different temperatures for the low-energy electron irradiated 4*H*-SiC sample.

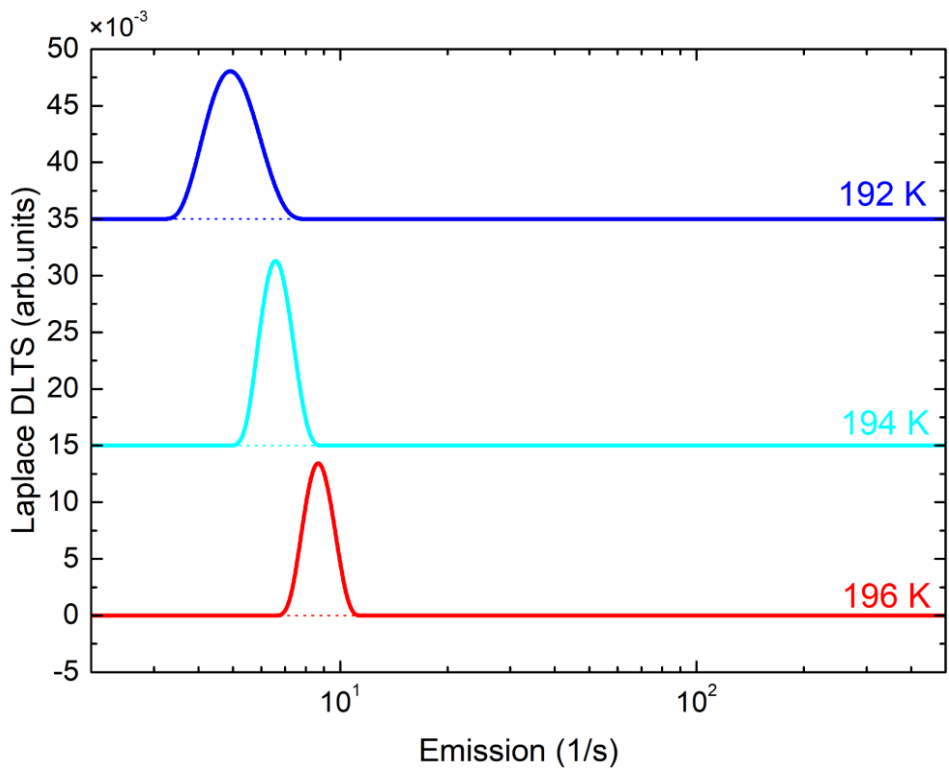

Figure 3. Laplace DLTS spectra measured at selected temperatures for the low-energy electron irradiated 4*H*-SiC.

Laplace DLTS measurements of EH1 reveal a single emission line (Figure 3). The estimated activation energy is 0.40 eV. As already proposed by Alfieri and Mihaela[3] EH$_1$ is assigned to C$_i$ related defect. Recently, Coutinho et al.[24] have assigned the metastable defect, known as

M-center[25, 26], to carbon interstitials. Combining the DFT calculations and isothermal DLTS measurements, they show that $M_1$ ($E_c$-0.42eV), one of four defects arising from the M-center, is $C_i$ located at the hexagonal lattice site, $C_i(h)$. Knezevic et al.[27] have applied DLTS and Laplace DLTS for studying the metastable defects in low-energy electron irradiated 4H-SiC and concluded that $M_1$ is in fact identical to $EH_1$. We, therefore, assign $EH_1$ (shown in Figure 3) to $C_i(h)$.

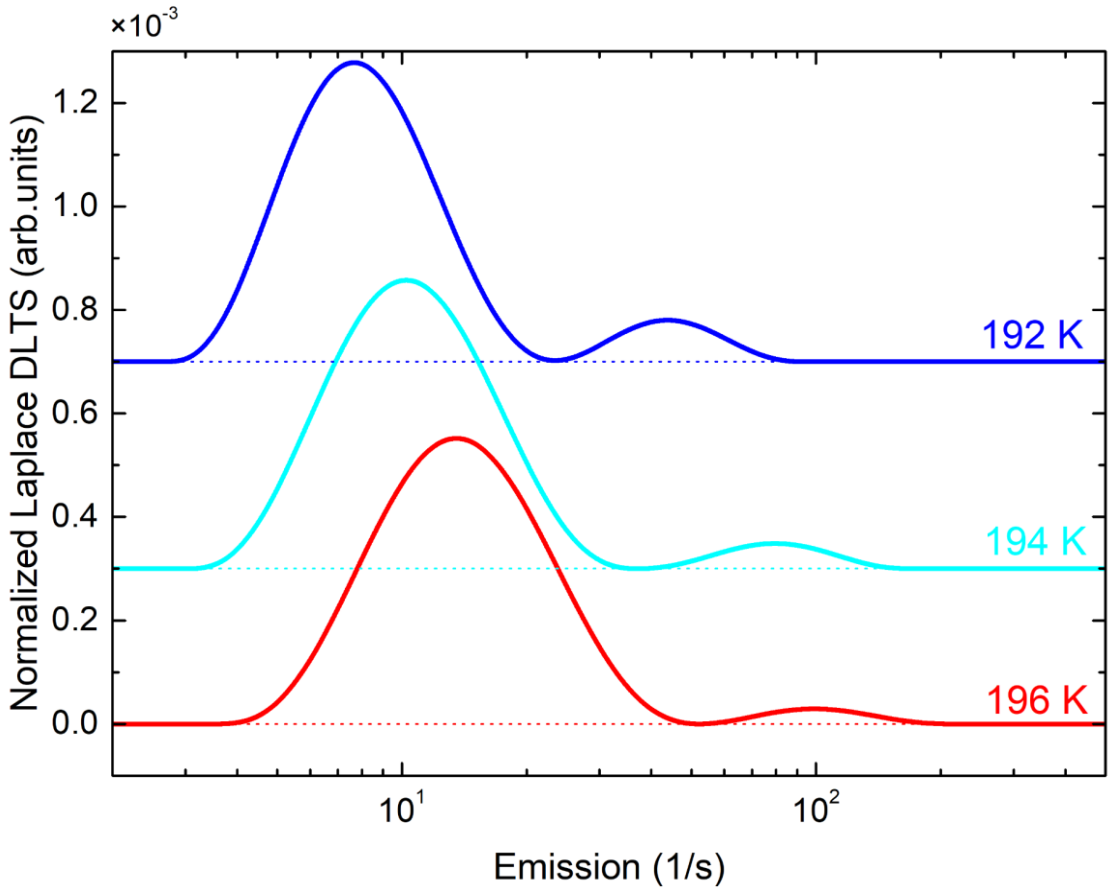

Figure 4. Laplace DLTS spectra measured at selected temperatures for the fast neutron irradiated 4*H*-SiC.

Figure 4 shows Laplace DLTS measurements for $S_1$ at different temperatures. Laplace DLTS measurements reveal that $S_1$ consists of two emission lines, with activation energies of 0.40 and 0.41 eV. They are assigned to $V_{Si}$ at -*h* and -*k* lattice sites. This is consistent with the previously published work done by Bathen et al.[12]

The difference between the results obtained by Bathen in the proton irradiated 4*H*-SiC samples, and those presented here (fast neutron irradiated) is in the ratio between two emission lines arising from the $S_1$. While in the proton irradiated samples the ratio is close to 1:1, in our case the ratio is around 1:6. The ratio shows the occupancy of the -*h* and -*k* lattice sites, and it indicates that the -*h* sites are more favorable than the -*k* sites (for the neutron-irradiated samples). The explanation for this difference is most likely due to the different sample preparations. In the proton irradiated study, the 4*H*-SiC samples were annealed upon irradiation, and thermal equilibrium between -*h* and -*k* sites was reached. In our case, no thermal annealing was performed upon the fast neutron irradiation.

Laplace DLTS has provided another indisputable evidence that $EH_1$ and $S_1$ originate from different defects. Moreover, it provides direct evidence that $EH_1$ consists of a single emission line arising from the $C_i$ at the -*h* lattice site, in contrast to $S_1$ which has two emission lines arising from the $V_{Si}$ at the -*h* and -*k* lattice sites. Therefore, Laplace DLTS can be used for easy and conclusive discrimination of the $EH_1$ and $S_1$ defects in n-type 4*H*-SiC material.

**ACKNOWLEDGEMENT**
This work was supported by the North Atlantic Treaty Organization Science for Peace and Security Program through Project No. G5674.